\documentclass[manuscript,screen]{acmart}

\usepackage{footnote}
\usepackage{makecell}
\usepackage{amsmath}
\usepackage{amstext}
\usepackage{subcaption}
\usepackage{xcolor}
\usepackage{hyperref}
\usepackage{algorithmic,algorithm}
\usepackage{booktabs, makecell, multirow}
\usepackage{float}

\newcommand{\dtoprule}{\specialrule{1pt}{0pt}{0.4pt}%
            \specialrule{0.3pt}{0pt}{\belowrulesep}%
            }
\newcommand{\dbottomrule}{\specialrule{0.3pt}{0pt}{0.4pt}%
            \specialrule{1pt}{0pt}{\belowrulesep}%
            }
\DeclareMathOperator{\arctantwo}{arctan2}

\AtBeginDocument{%
  \providecommand\BibTeX{{%
    \normalfont B\kern-0.5em{\scshape i\kern-0.25em b}\kern-0.8em\TeX}}}

\setcopyright{acmcopyright}
\copyrightyear{2021}
\acmYear{2021}

\acmBooktitle{Proc. ACM Interact. Mob. Wearable Ubiquitous Technol.}
\acmISBN{978-1-4503-XXXX-X/18/06}

\begin{document}

\title{A Novel IoT-Based System for Ten Pin Bowling}

\author{Ilias Zosimadis}
\authornote{Both authors contributed equally to this research.}
\email{zosimadis@csd.auth.gr}
\orcid{0000-0001-6352-9962}
\affiliation{%
  \institution{Aristotle University of Thessaloniki}
  \city{Thessaloniki}
  \country{Greece}
  \postcode{541-24}
}

\author{Ioannis Stamelos}
\orcid{0000-0001-9440-3633}
\affiliation{%
  \institution{Aristotle University of Thessaloniki}
  \city{Thessaloniki}
  \country{Greece}}
\email{stamelos@csd.auth.gr}

\begin{abstract}
  Bowling is a target sport that is popular among all age groups with professionals and amateur players. Delivering an accurate and consistent bowling throw into the lane requires the incorporation of motion techniques. Consequently, this research presents a novel IoT-Cloud based system for providing real-time monitoring and coaching services to bowling athletes. The system includes two inertial measurement units (IMUs) sensors for capturing motion data, a mobile application and a cloud server for processing the data.
  First, the quality of each phase of a throw is assessed using a Dynamic Time Wrapping (DTW) based algorithm. Second, an on device-level technique is proposed to identify common bowling errors. Finally, an SVM classification model is employed for assessing the skill level of bowler athletes. We recruited nine right-handed bowlers to perform 50 throws wearing the two sensors and using the proposed system. The results of our experiments suggest that the proposed system can effectively and efficiently assess the quality of the throw, detect common bowling errors and classify the skill level of the bowler.  
\end{abstract}


\maketitle

\section{Introduction}

The emerging era of Internet of Things (IoT) has revolutionized the sports industry by providing monitoring tools and mechanisms for sports performance analysis. IoT enabled solutions coupled with wearable technologies, machine learning, and big data techniques allow athletes and coaches to estimate performance capabilities, measure efficiency, and develop effective training strategies. Due to the plethora of opportunities for IoT in the smart-sports industry, sports performance analysis systems have attracted considerable research attention. Sports such as basketball, tennis, swimming, skiing, and rowing are some of the activities researched for performance analysis with IoT technologies \cite{9058658}, \cite{9174823}. One activity that has not received much attention from researchers is the ten-pin bowling. Ten-pin is one of the main five bowling sports variations, namely nine-pin, candle-pin, duck-pin, and five-pin bowling. The goal of this activity is to knock down ten pins (which are positioned at the end of an alley) using a bowling ball. It can be played by amateurs or professionals at almost all age groups and all genders. These facts make bowling one of the most popular target sports in the world. 

Although the objective of ten-pin bowling seems an easy task, bowlers need to incorporate fundamental concepts and techniques into their games in order to deliver a good performance. Five and four-step approaches are the two main techniques that are used by the bowlers. The two techniques define rather accurately the motion of a bowler during a ball delivery and facilitating them to perform an accurate and consistent throw. The synchronization between the swing arm and the gait of a bowler is the main contribution of the two approaches for delivering an accurate throw. Fundamental concepts such as flexibility, balance, timing, and stamina are common to all bowlers. It has been evidenced that professional and semi-professional bowlers have unique anthropocentric and strength characteristics compared to non-bowlers \cite{Anthropometric}. These characteristics are developed through the practice of efficient movement techniques. To achieve better performance, bowlers and coaches need to monitor these motion techniques during the training process in order to assess the performance, identify, and correct errors. 

Despite the importance of monitoring and reporting the quality of motion in bowling, there are few tools available for athletes and coaches. The existing systems depend on infrastructure-based technologies such as LiDAR sensors \cite{specto} and high-frame rate cameras \cite{itarc}. They mostly provide feedback about the trajectory of the ball into the lane, omitting motion technique-related information. In addition, these systems are expensive and unaffordable for individual bowlers. Hence, these systems are not suitable for monitoring and assessing the goodness of bowlers motion techniques.

In this work, we propose a novel IoT-based system designed to monitor and assess bowlers' training performance using commercial inertial measurement units (IMUs) and Cloud technology. Two sensors are positioned on the bowler's wrist and leg and stream data through Bluetooth Low Energy (BLE) to a mobile device. The mobile device utilizes the sensors' data to provide real-time feedback for the quality of throws and detect errors related to motion techniques. Users can upload and share their performance with other users through a cloud server. Moreover, we use a machine learning model to distinguish the skill levels of bowlers, between expert, intermediate, and novice. The proposed machine learning model is deployed on the cloud server. Our experiments show that the proposed system can effectively detect errors in bowlers' motion techniques, reaching a precision of 90\% and recall 84\%. We also proved that our system efficiently assesses the quality of throws and characterizes the skill level of bowlers (87\% prediction accuracy). To summarize, the main contributions of this study are as follows:

\begin{enumerate}
    \item A novel IoT and Cloud based system for assessing and monitoring the motion performance of bowlers. In comparison with existing commercial solutions, the proposed system does not depend on infrastructure-based technologies and, to the best of our knowledge, is the first that uses bowler's motion characteristics to assess the motion performance and to provide real-time feedback. 
    \item An efficient on device-level technique for identifying common bowling errors and assessing the motion quality of bowlers.   
    \item An SVM based model, deployed on a cloud infrastructure for classifying the skill levels of players.
\end{enumerate}

The paper is organized as follows. Section II presents an overview of the related literature. Section III describes motion techniques and common errors for a bowling throw. Section V presents the workflow of the proposed system, which includes data preprocessing, quality assessment, error detection and skill assessment. Section VI reports the experimental results and Section VII concludes this study.  The technical details of the proposed system, including the system hardware, the mobile application development, and the cloud architecture are discussed on the complementary article.

\section{Related Work}

Section \ref{Wearable in Sports} shows how other researchers applied wearable computing in other sports and physical activities, such as baseball, ski and swimming. Section \ref{Ten-pin research} shows studies that have examined the use of technologies in ten-pin bowling.

\subsection{Wearable computing in Sports}\label{Wearable in Sports}

Monitoring and assessing athletes' performance for arm movement-related activities such as tennis, cricket, golf, and strength training have been an area of research in many recent studies \cite{tabrizi2020comparative}, \cite{kos2018tennis}, \cite{gawsalyan2017upper}, \cite{kim2017golf}, \cite{jia2019mobile}, \cite{9046259}. Several studies have also examined the usage of IMU sensors in sports. For example, the authors in \cite{10.1145/1620545.1620578} introduced the concept of a wearable assistant for swimmers. Using acceleration sensors they monitor a wide range of swim parameters, including the swimming velocity, the time per lane and the number of strokes. These parameters and swim style-specific factors, like body balance and body rotation, are used to monitor swim performance and provide feedback to the athlete. In \cite{khan2020generalized}, the authors proposed a new framework for skill assessment activity recognition and motion detection. Raw motion data (3-axis acceleration) collected with IMUs are used to train an SVM classifier for movement classification and activity recognition. In addition, a second SVM classifier was trained to provide confidence scores for skill assessment. The effectiveness of the aforementioned classifiers were examined in two case studies: (1) automated scoring in gymnastics and (2) surgical skill assessment. The achieved results showed that their proposed framework has high assessment accuracy. In \cite{10.1145/3478076}, the authors proposed a metric called LAX-Score to quantify the athletic performance of a lacrosse team. The proposed metric was statistically validated on IMU and biometric data obtained from a lacrosse team over the course of a season. They investigated the correlation between the extracted data features (obtained with the sensors) and the proposed performance metric. Their analysis suggested that some features have greater impact on athlete's performance than others. The system in \cite{10.1145/2968219.2968535} uses accelerometer and gyroscope data to train a classifier and recognize left and right alpine ski turns. The authors in \cite{velloso2011towards} developed a weight lifting assistant that uses motion sensors mounted on athlete's body and provides qualitative feedback on the user's performance. The authors in \cite{derungs2018regression} proposed a mistake-driven skill estimation approach in Nordic Walking using IMUs. The presented approach assess three mistakes related to athlete's stride. A Bayesian Ridge Regression model was trained for each mistake type and used to estimate the movement skill of the athlete. The skill estimation approach was evaluated using 10 Nordic Walking beginners with their results suggesting that all mistakes can be estimated an normalised RMSE of 24.15\%.

\subsection{Ten-pin bowling research}\label{Ten-pin research}

A few studies have examined the use of technologies in bowling sports. Hung \emph{et al.} \cite{han2010measuring} developed an artificial-neural-network model for predicting the bowlers' knee forces during a throw. Investigating and monitoring the forces of lower limbs can help to prevent injuries. Joint forces relating to the lower limbs motion were estimated from Euler angles using inverse dynamics methods. They trained an ANN model to predict knee forces using Euler angles obtained from a high speed camera. The ground truth data were obtained using two force plates and the accuracy of the model was determined by the coefficient of determination which was found to be high for all experiments in the study. However, this system cannot assess the quality of bowlers' technique as it only relies on lower limb joint forces.

King \emph{et al.} \cite{king2011bowling} presented a miniature wireless inertial measurement unit embedded in a bowling ball that can be used to deduce the dynamics of the ball. Acceleration and angular velocity data were wirelessly   transmitted by the IMU during the bowler throw. The collected data was used to calculate the spin dynamics of the ball, which are essential for measuring the performance of the release phase. Unlike our study, arm swing motion and gait characteristics were not taken into account in this type of system, which means that it cannot measure the complete motion performance. Moreover, the use of sensors within a bowl raises several practical concerns, like the conformance of the bowl with standard official match rules. 

Tung Mun Hon \cite{Tung} developed an interactive graphical user interface LabVIEW program, which can be used to visualize data related to the arm swing motion. The data were obtained from an IMU sensor, which was mounted on the bowlers' wrist. Angular velocity and roll Euler angles were used to derive the arm swing speed, time, and wrist pronation parameters. The three parameters were displayed on a LabVIEW program. However, this system leverages only a limited number of motion parameters, hence it does not provide a comprehensive assessment of bowlers' techniques. Besides, using Euler angles to measure the pronation of the arm could cause the gimbal-lock problem.   

Finally, to the best of our knowledge, the proposed system is the first IoT-based approach in bowling sports that can assess the quality of a throw, identify motion errors and characterize the skill level of bowlers.

\section{Investigated technique}

\begin{figure}[!t]
\centering
\includegraphics[height=1.5in, width=5.25in]{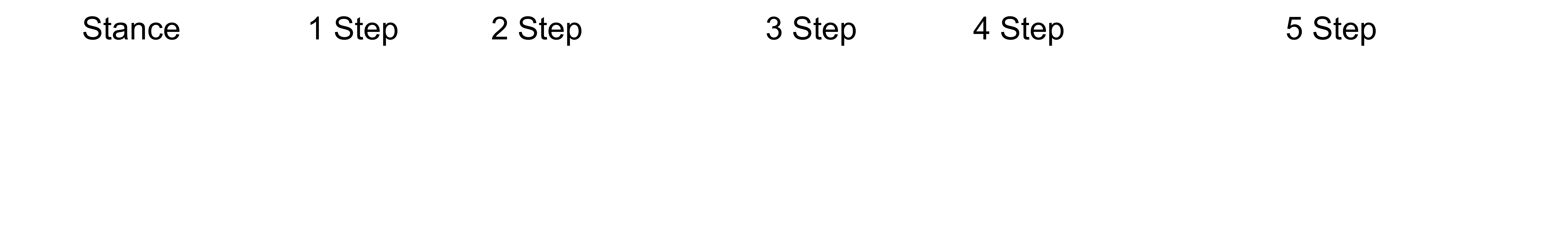}
\caption{The five step approach performed by an expert bowling player.}
\label{fig_approach}
\end{figure}

In this study, we focus on one of the most popular bowling movement techniques, namely the five step-approach, which can be performed by bowlers of all skill levels for a smooth delivery. Figure \ref{fig_approach} depicts the phases of the five-step approach. Initially, the bowler holds the ball near the horizontal plane and aims it into the alley. It must be noted that the pendulum-arm involves anterior, posterior movements in the vertical plane and pronation/supination of the wrist. As the name of the technique suggests, the bowler performs five normal steps with a simultaneous motion of the dominant arm (the arm that holds the ball). The synchronized motion sequence of the legs and the pendulum-arm plays a significant role in a successful ball delivery \cite{wiedman2015bowling}. Although this motion sequence is unique for each bowler and derived from their style, there are some key elements that are common to all athletes. The swing motion of the arm during the  third and fifth steps (gait with the non-bowling leg) is independent of the bowlers' style and is the main element that affects the timing and accuracy of the throw. It is a common practice among coaches to evaluate the quality of a throw by using the position of the arm during these phases. For instance, a wrong arm swing during the third step could indicate a problem with bowler's timing. Thus, in our approach, to assess the quality of a throw, we focus on arm swing motion during the synchronized movement of the non-bowling leg.

There are many errors that may happen during a bowling throw \cite{grinfelds2011right} (see section 5). The identification of these errors is an important task for bowlers and coaches in order to achieve good performance. However, due to various parameters associated with bowlers' motion technique, it is infeasible to examine every possible error. Thus, we select four errors in our study which are described below:
\begin{itemize}
    \item \emph{Error 1}: The gait speed of the non-bowling foot is too fast or too slow. On the five-step approach, the speed of the non-bowling foot is important for delivering an accurate throw. Bowlers with wrong gait speed tend to have a lack of fluid motion and erratic ball delivery.
    \item \emph{Error 2}: The backswing (the maximum angle of arm swing) is too high or too low. A high backswing may lead to a ball delivery with too much speed and possibly an inaccurate throw. Conversely, a low backswing may lead to a forced release of the ball due to the low speed of arm swing.
    \item \emph{Error 3}: The overall delivery is too fast or too slow. The arm swing speed and the synchronized motion of the legs are significant factors for an accurate ball delivery. A fast or slow ball delivery may lead to a throw with poor timing or inconsistent release. A common practice to identify this error is to measure the position of the swing arm during the last step in relation to the non-bowling foot when the latter stops moving. If the position of the swing arm is behind the stationary non-bowling foot, the bowler performed a throw with late timing. Conversely, if the position of the swing arm is in front of the non-bowling foot, then the bowler performed a throw with early timing. A throw with late or early timing may indicate a delivery that was performed too fast and too slow, respectively.
    \item \emph{Error 4}: The ball path and direction of the throw is inconsistent. This error is often caused by unfavorable positioning of the ball in the stance phase. Performing such throws may also lead to inconsistent timing on the early steps of the bowler. Similar to \emph{Error 3}, coaches measure the position of swing arm during the third step in relation to the non-bowling foot in order to classify the throw's timing. If the swing arm is positioned in front of the non-bowling foot when the third step completes (the non-bowling foot is in contact with the ground), then the first steps of a throw have timing issues. 
\end{itemize}

These four errors are specifically selected and verified by our bowling experts because they are commonly observed, and may be the manifestation of an athlete’s wrong approach to throwing. Removing these errors from one’s move assures a much better technique and overall performance.

\section{Methods}

\begin{figure*}[t]
\centering
\includegraphics[scale=0.7]{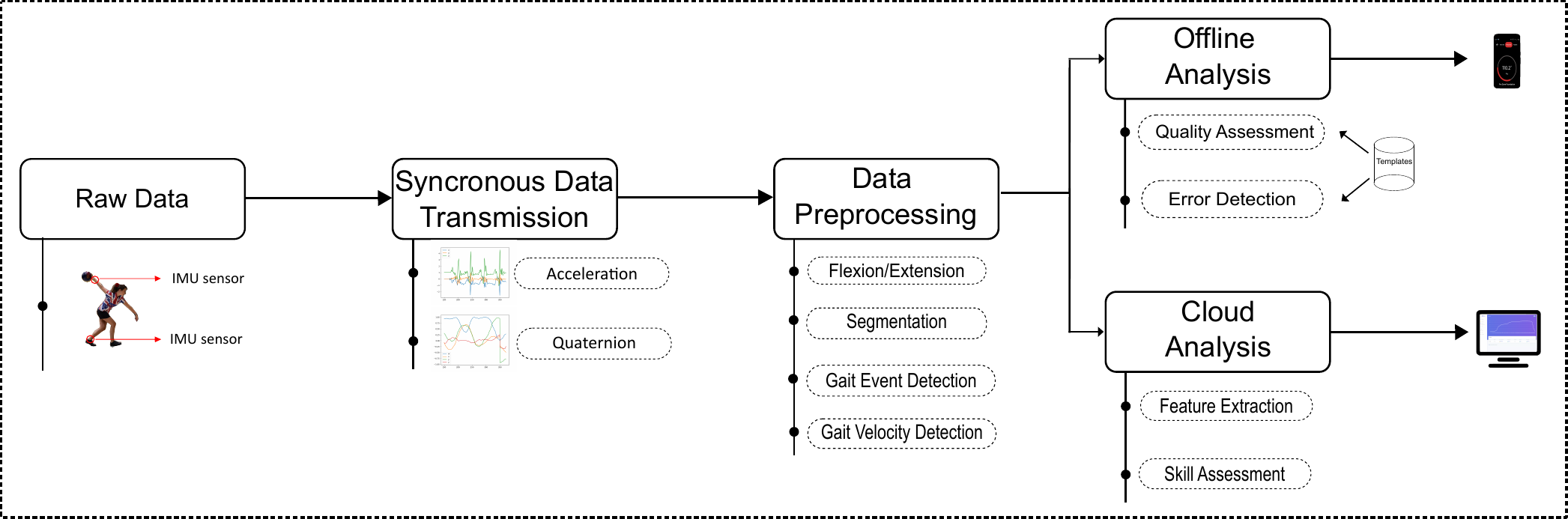}
\caption{The data processing flow of the proposed system.}
\label{fig_workflow}
\end{figure*}

Figure \ref{fig_workflow} presents an overview of the proposed workflow of our system. The workflow is comprised mainly of three stages, which are (i) data preprocessing, (ii) quality assessment, error detection and (iii) skill assessment. After the transmission of the synchronized data to the mobile device, data preprocessing is performed to extract the swing angle, segment the data, and detect gait events and velocity. The system assesses the quality of the throw and detects common bowling errors, using the synchronized swing angle and gait data events for the corresponding segment. Similarly, the system assesses the skill level of the bowler using a supervised machine learning workflow. The following subsections provide details about the workflow of the proposed system.     

\subsection{Data Preprocessing}

\subsubsection{Swing Angle and Segmentation} 
\begin{figure*}[!t]

\subfloat[]{\includegraphics[width=2.5in]{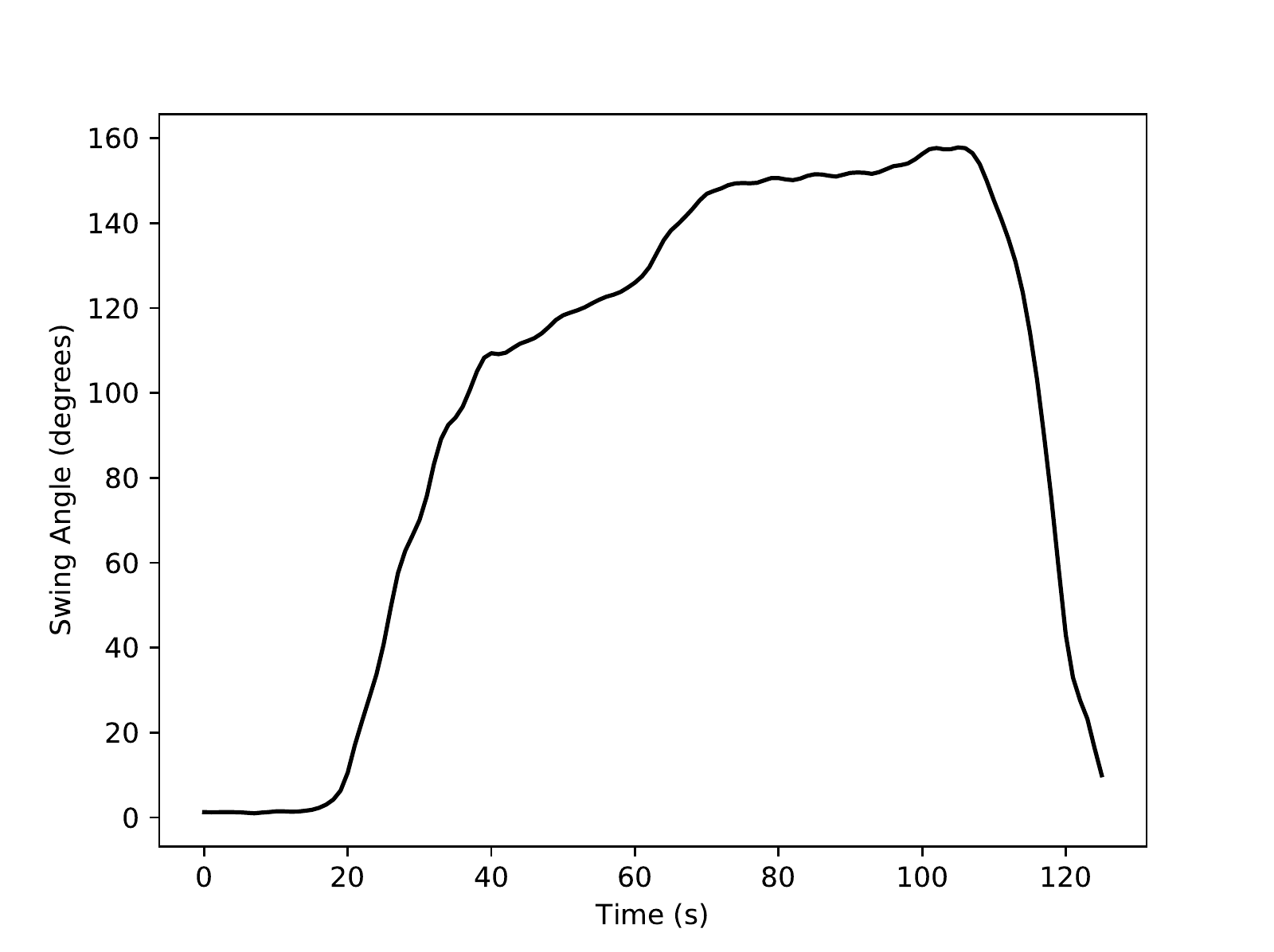}\label{fig_swing_angle}} 
\subfloat[]{\includegraphics[width=2.5in]{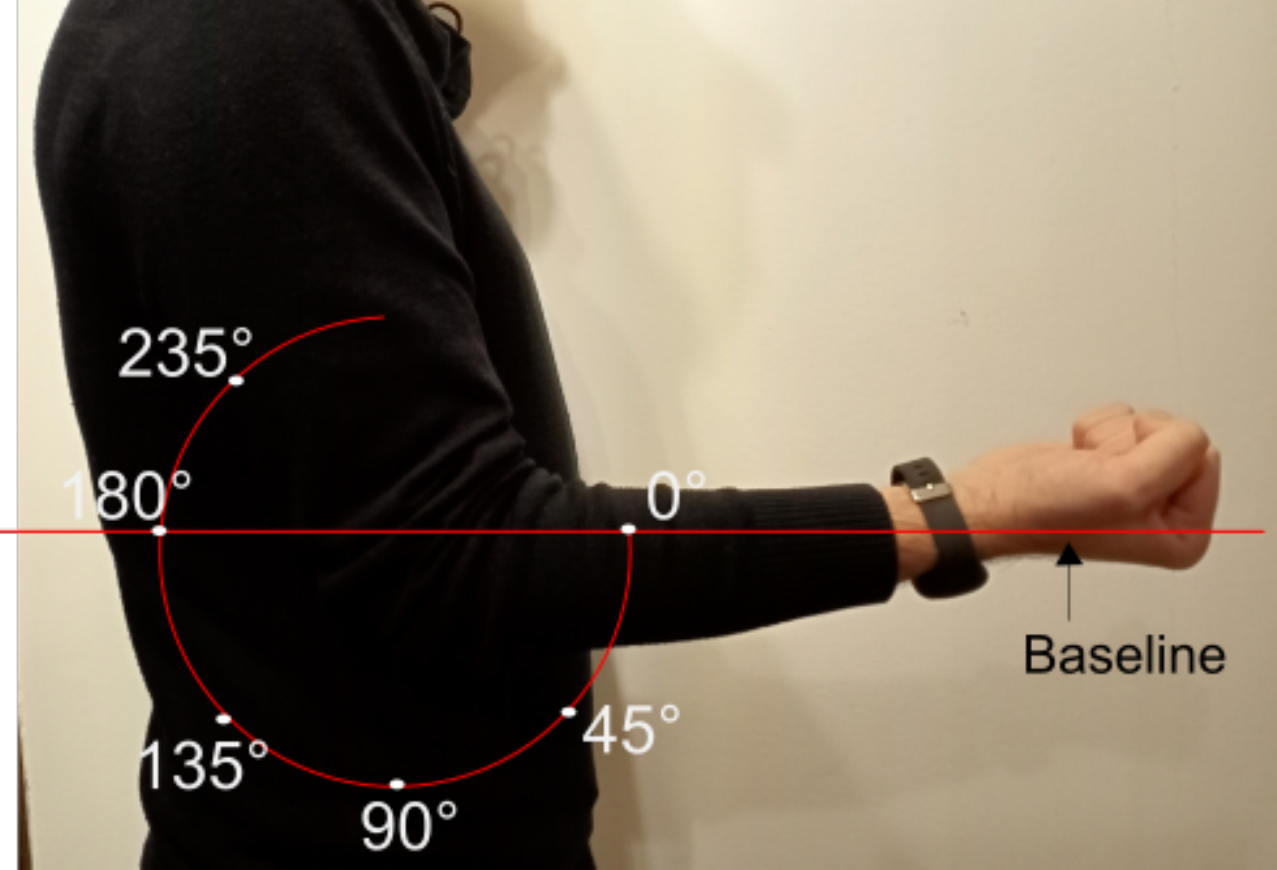}\label{fig_baseline}}
\caption{The swing angle of a bowler (a) and (b) the baseline with the corresponding swing angles from that baseline.}
\label{fig_hand}
\end{figure*}

As we show in Section II, a bowler's arm swing is a composite motion that involves flexion/extension of elbow and shoulder and pronation/supination of wrist. We define the flexion/extension angle between the wrist and the transverse plane as the bowler's swing angle (Figure \ref{fig_arch}). In our system, the baseline acquired during the game setup (section II.B) is acting as a reference point for swing angle computation. The bowler's arm motion during a throw will result in positive swing angles typically in the range of $0^\circ$ to $250^\circ$ (see figure \ref{fig_baseline}). 

To estimate the orientation of the sensor and consequently the swing angle we used quaternions. The quaternions are a four-dimensional number system that extends complex numbers and can be used to encode rotations in a 3D coordinate system. Unlike Euler angles, quaternions do not suffer from gimbal lock \cite{johnson2003exploiting}. We obtain four-dimensional quaternion data $q_{t} = [q_{0}, q_{1}, q_{2}, q_{3}]$ with respect to time $t$, from the IMU wrist sensor. There are three coordinate frames in respect to swing analysis: the inertial frame $i$, the sensor frame $s$ and the baseline frame $b$. The inertial frame is a stationary frame, having the origin at the center of the earth and its axis aligned with respect to the stars. The sensor frame is the coordinate frame of the moving IMU sensor on wrist. This frame determines all of the inertial measurements of the sensor. The baseline frame is a local stationary frame, in which the axis are fixed with respect to the axis of the sensor at baseline position. In our analysis, we are interested in finding the position and orientation of the sensor frame relative to the baseline frame. The coordinate frames for the swing analysis are shown in Figure \ref{fig_sensor_hand}. We denote all of the unit quaternions, which are obtained by the sensor and  quantify rotation from the frame $i$ to frame $s$ as  ${}^{s}_{i}q = [q_{0}, q_{1}, q_{2}, q_{3}]^T$. The equation to produce unit quaternions from the data acquired by the sensor is:

\begin{equation}
     Q_{u} = \frac{q}{\|q\|} \text{ where } \|q\| = \sqrt{q_{0}^2 + q_{1}^2 + q_{2}^2 + q_{3}^2 }
\end{equation}

To compute the swing angle using quaternion operations, we first split the rotations of the arm into swing and twist components. The swing-twist decomposition \cite{grassia1998practical} is used to determine the two components of rotation. By defining the sensor's y-axis as the twist axis, it results in a twist quaternion $q_{twist}$ consisting of rotations around this axis and a swing quaternion $q_{swing}$ with the remaining rotation around the horizontal axis. The $q_{swing}$ orientation quaternion of frame $s$ with respect to frame $b$ can be expressed by the following equation:  
\begin{equation} \label{eq:2}
     {}^{b}_{s}q(t) = q_{baseline} \otimes q_{swing}^{*}(t) 
\end{equation}
where $q^{*}$ is the conjugate quaternion and  $\otimes$ is the Hamilton product between the two quaternions \cite{hamilton1866elements}. Let ${}^{s}u = [0,1,0]^{T}$ be a unit vector expressed with respect to the frame $s$. We want to rotate the vector ${}^{s}u$ in respect to the frame $b$ by applying the ${}^{b}_{s}q$ orientation quaternion. According to \cite{kuipers1999quaternions}, unit quaternions can be applied to rotate a vector $u \in \mathbf{R}^{3}$ by representing it as pure quaternion. The pure quaternion of the vector ${}^{s}u$ defined as: 
\begin{equation} \label{eq:3}
     {}^{s}u_{q} = [0,{}^{s}u]^{T} = [0,u_{1},u_{2},u_{3}]^{T} 
\end{equation}
Then, we use the orientation quaternion from equation \eqref{eq:2} to rotate quaternion ${}^{s}u_{q}$ with respect the frame $b$:
\begin{equation} \label{eq:4}
     {}^{b}u_{q}(t) = {}^{b}_{s}q(t) \otimes {}^{s}u_{q} \otimes {}^{b}_{s}q^{*}(t)
\end{equation}
where ${}^{b}u_{q}$ is another pure quaternion, namely a vector rotated by the quaternion ${}^{b}_{s}q$. We define the vector part of ${}^{b}u_{q}$ as $v$. By using $v$ and the vector ${}^{s}u$ we can find the swing angle $\omega$ as:
\begin{equation} \label{eq:5}
     \omega = \arctantwo (\|v \times u \|,v \bullet u) 
\end{equation}
where $\omega$ is the swing angle defined in the range $(-\pi,\pi]$, which can be mapped to $[0,2\pi) $ by adding $2\pi$ to the negative values.

Finally, the swing angle estimation $\omega$ is used to detect the start and end points of the bowler's throw data stream. To identify the segments of the swing angle automatically and in real-time, we used a varying-time window-based technique described in \cite{OKEYO2014155}. Figure \ref{fig_swing_angle} displays the swing angle of a bowler. 

\subsubsection{Gait Events and Velocity}

\begin{figure*}[!t]
\centering
\subfloat[]{\includegraphics[width=2.5in]{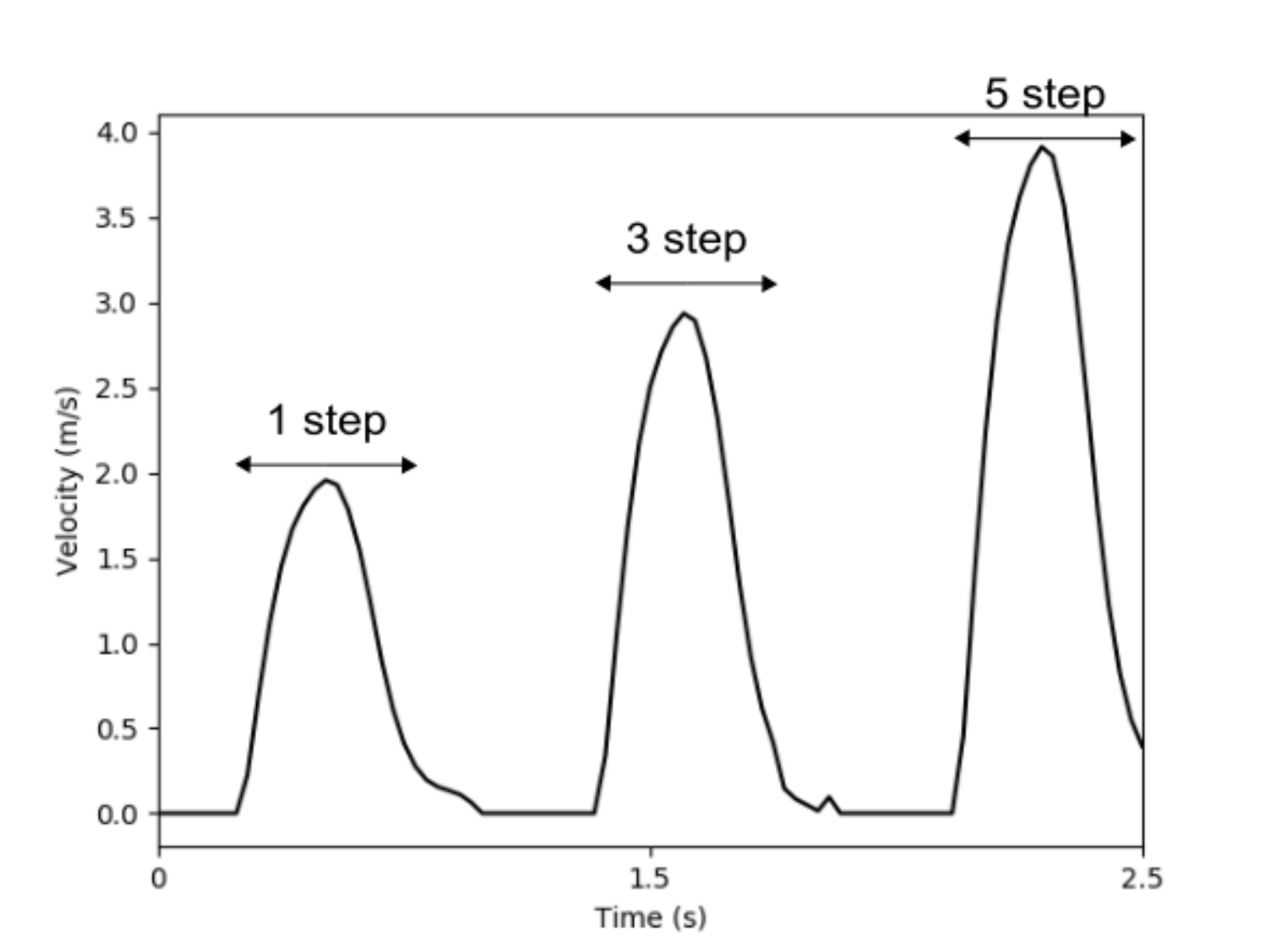}\label{fig_velocity}} \\
\subfloat[]{\includegraphics[width=2.5in]{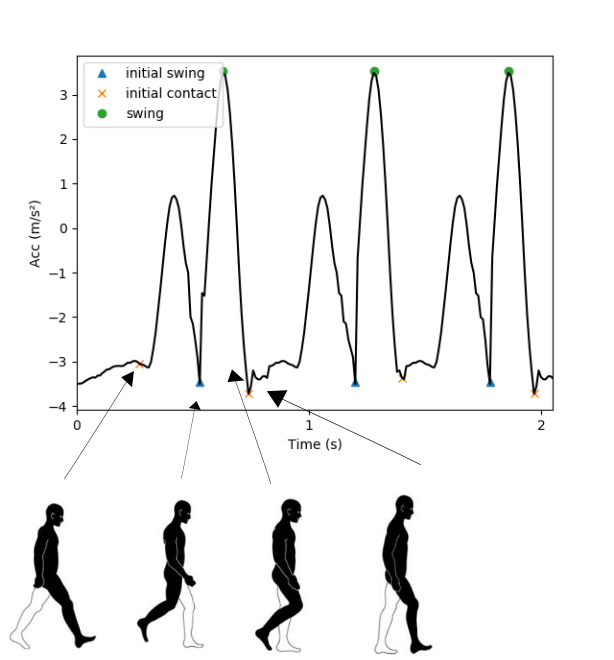}\label{fig_gait}}
\caption{The velocity of the non-bolwing leg (a) and the filtered signal (b)  with the corresponding gait phases during a throw. }
\label{fig_graphs}
\end{figure*}

To develop our quality and skill assessment algorithms we first have to extract bowler's gait characteristics. Human gait cycle can be defined as a repetitive pattern involving the forward movement of the lower two extremities \cite{wang2012walking}. A normal forward step has two phases: a swing and a stance. Stance is the period between the initial contact of the foot on the ground and the final lift of toes off the ground. Conversely, swing phase is the period amid which the foot is not touching the ground. During this phase the foot rapidly accelerates forward followed by a deceleration to the initial contact of a stance event. Figure \ref{fig_gait} depicts the phases of a normal human gait cycle.

As we show in section II, to deliver the ball into the lane, bowlers have to perform five or four normal steps. Besides, the gait characteristics of the non-bowling foot can provide valuable insights into the quality of the throw. Thus, to identify gait phases, we use a stride detection technique based on local minimal prominence characteristics of strides associated with the magnitude of accelerometer data \cite{s18020676}. This method can identify the start and the end of a gait cycle, the start of a stance event and swing event. Figure \ref{fig_gait} shows the result of this method applied to the leg with the IMU sensor, during a five-step approach. In the initial and final position of the throw, the non-bowling foot is stationary on the ground, having zero acceleration. There are three strides detected for the non-bowling foot, while for each stride we identify the initial contact of the stance phase (red circle on the figure) and the pre-swing event of the swing phase (blue circle on the figure). The highest value of acceleration for each stride corresponds to the mid-swing event of the swing phase. Using the gait events detected by this method, we estimated: number of strides, swing period, stance period, ratio of swing and stance. The swing period is the interval between the initial swing and the initial contact phases. The stance period is the interval between the initial contact and the pre-swing (of the next stride) phase. Typically, the ratio between the stance and swing phase is $60\%$ and $40\%$, respectively \cite{iosa2013golden}.

Another significant factor for the quality of the throw is the velocity of the bowler's steps. To calculate velocity for each stride of the non-bowling leg, we used trapezoidal integration method of accelerometer data. However, due to noise, direct integration of accelerometer can generate significant amount of drift error \cite{THONG200473}\cite{han2010measuring}. To address this issue we use the velocity estimation method proposed in \cite{s18020676}. The magnitude of acceleration is passed through a second-order high pass Butterworth filter with a 1Hz cutoff frequency and 50Hz sampling rate and a filtfilt filter. The filtfilt filter is used to correct the phase distortion introduced by the second-order high pass Butterworth filter. Then, for each stride of the filtered total acceleration, we measure the gait velocity using the trapezoidal integration method. The gait velocity obtained using this method is shown in Figure \ref{fig_velocity} .

\subsection{Quality Assessment and Error Detection}

\begin{algorithm} 
    \caption{: The quality assessment algorithm.} 
    \label{alg1} 
    \begin{algorithmic}[1] 
        \renewcommand{\algorithmicrequire}{\textbf{Input:}}
        \renewcommand{\algorithmicensure}{\textbf{Output:}}
        \REQUIRE Swing Angle Templates: {$T^{k}_{i} = [A^{k}_{1,i},A^{k}_{2,i},...,A^{k}_{N,i}]$ $i=1,2,...,S$ $k=1,2,...,K$};
        
        Swing Angle of User: {$U^{k} = [A^{k}_{1}, A^{k}_{2},...,A^{k}_{N}]$};
        
        // S, N, K is the number of templates, length of data and number of strides respectively. 
        \ENSURE  A set of quality degrees $QD$ of size K, between user's strides and templates. 
        
        \STATE Let $QD=[1..K]$;
        \FOR{each stride $k=1$ to K}
            \STATE Let $Distance=[1..S]$;
            \FOR{each template i=1 to S}
                \STATE $Distance[i] \mathrel{+}= DTW(T^{k}_{i},U^{k})$;
            \ENDFOR
            \STATE $QD[k]=(1 - \frac{MIN(Distance)}{MEAN(Distance^{*})})\times 100$
        \ENDFOR
        \RETURN $QD$
    \end{algorithmic}
\end{algorithm}

After the synchronous data acquisition, we use the extracted information from swing and gait analysis to detect common bowling errors and assess the quality of a throw. The process of identifying the common errors and the throw quality is executed on the user's mobile device after each throw. An example of the results obtained by this analyses is shown in Figure \ref{fig_screen_overview}.

\subsubsection{Quality Assessment of throws}

In section II, we noted that the proper positioning of the arm during the steps of the non-bowling leg plays a significant role in a successful ball delivery. Thus, an approach based on the comparison with a throw considered to be close to ideal was employed in the quality assessment of the throw using the extracted swing angle and gait events of the non-bowling leg. This process aims to measure how well each phase of the throw is performed and provides visual feedback of the athlete's performance on a mobile device.

Our approach to assessing the quality of each phase of the bowler's throw is to use template sample data obtained from the coach and measure the similarity between the bowler's swing angle and the corresponding templates. These templates are motion data of throws performed by a coach, representing the optimal way to deliver the ball into the alley using a particular style (e.g. five-step approach). In a similar manner to user motion data, swing and gait analysis are used to extract angle and gait events for the templates. The extracted information from the templates of the user's corresponding style is pre-stored in the mobile device. There are many ways to measure the similarity between two time series \cite{keogh2003need}. In our quality assessment process, we apply the dynamic time warping (DTW) algorithm, which can measure the similarity between two time series after finding the optimal alignment between them. Unlike other metrics such as Euclidean Distance, DTW can accurately measure the distance of two similar data series even if they are out of phase in the time domain or have unequal length. Consequently, due to the variation of the speed of collected motion data, the properties of DTW method make it well suited for our system. Despite the ubiquitous use of the DTW in pattern recognition studies, this technique requires a high computational complexity and therefore affecting the performance of the application. Notably, the time complexity of this algorithm is $O(NM)$, where $N$ and $M$ are the lengths of the two data series which are used as input to measure the distance. To address this issue, we choose only a small number of templates to be stored and processed for each athlete. In this way, we reduce the total number of comparisons and, therefore, the overall time delay of the process. The impact of the number of templates in our approach will further be examined in Section IV. The DTW measures a distance quantity, meaning that higher values indicate a weaker match between the two motions. To quantify the quality of an athlete's motion and provide a more intuitive interpretation of the DTW similarity, we use the $Quality Degree$ (QD) \cite{9046259} defined as: 

\begin{equation} \label{eq:6}
     QD(x) =\frac{AD - MD}{AD} \times 100
\end{equation}

where $MD$ is the minimum DTW distance of the athlete's motion data $x$ and the applicable template data, and $AD$ is the average DTW distance between $x$ and the remaining incompatible templates. The range of the $QualityDegree$ is set to $[0, 100]$. An athlete's motion, which is similar to the corresponding optimal motion of his coach, is considered to have a high $QualityDegree$. In order to measure the quality for each significant phase of the throw, we use the swing angle and gait events extracted from the synchronized sensor data. Using the gait events of the non-bowling leg, we partitioned the swing angle data into segments defined by the start and end of the strides. For each of these segments and the corresponding templates, we calculated the DTW distance. Subsequently, we used the DTW distances to estimate the $Quality Degree$ of each segment. The process of throw quality assessment is summarized in Algorithm \ref{alg1}.

\subsubsection{Common Error Detection}

We developed a method to detect the four common bowling errors described in section II and give visual feedback for error conditions to the athlete. This method compares key statistical parameters between athletes and templates data. The parameters are defined by the coach and are pre-stored in the athlete's mobile device. To detect \emph{Error 1} on the bowler throw, we use the velocity data of the non-bowling leg, obtained by the gait analysis. Let $u^{k}_{avg}$ be the average velocity for stride $k$ and $v^{k}_{avg,i}$ be the average velocity of template $i$ for stride $k$. Then, if at least one of the $k$ deviations $D^{k}_{1} = | u^{k}_{avg} - m(v^{k}_{avg})| $, where $m(v^{k}_{avg})$ is the \emph{mean} of averaged templates velocities for stride $k$, surpasses a predefined range $[0,\epsilon]$, an \emph{Error 1} is detected. Similarly, to identify \emph{Error 2} we use the maximum swing angle of the throw and the averaged maximum swing angle of the templates. When the deviation $D_{2} = | A_{max} - m(A_{max,i})|$ exceeds a certain user defined range, an \emph{Error 2} is detected. 

\emph{Errors 3} and \emph{4} are highly dependent on the position of the arm during gait events of the non-bowling leg. Thus, the synchronized data of swing angles and gait events are used to identify the arm position. Bowler's arm angle on the initial contact gait event in the last stride (the fifth step) , is used to detect \emph{Error 3}. If the deviation $D_{3} =|A_{contact} - m(A_{contact,i})|$ exceeds the predefined range of $[0,\epsilon]$, the system alerts the users for \emph{Error 3}. The $A_{contact}$ is the swing angle during the initial contact event of the non-bowling foot in the last stride and $m(A_{contact,i})$ is the corresponding averaged swing angles of the templates. The last error that our system can report is \emph{Error 4}, which is determined by the position of the arm during the initial contact of non-bowling leg in the third stride. Thus, for throws with deviations $D_{4}=|A_{contact} - m(A_{contact,i})|$ which are out of the corresponding predefined range, we detect \emph{Error 4}. 

As we showed, our approach can detect each type of error using a range $[0,\epsilon_{i}]$. The upper bounds of the range, namely $\epsilon_{i}$, are defined by the coaches and are prestored on athletes' mobile devices.

\subsection{Skill Assessment}

A machine learning approach is employed, to assess the skill of a bowler. Our approach includes data preprocessing, feature extraction and classification. The training, evaluation, and hyper-parameter tuning of the model were performed locally, and the resulted model was deployed on the cloud. After the end of a game, the bowlers can upload their data on the cloud to perform the skill assessment analysis and visualize their performance. These data consist of the swing and gait analysis results which are extracted during the game on athlete's mobile device. There are three different skill levels - novice, intermediate and expert - and each throw of the game is labeled with the corresponding level by the skill assessment analysis.

In the data preprocessing step, we removed the throws which were not related to any particular bowling style (e.g. 5-step approach). Noise removal was applied by swing and gait analysis, therefore no additional filters were used in the preprocessing step. For each throw, we extracted 21 statistical features from swing angle, gait acceleration, and gait velocity  (7 features x 3 axis). These features are then used to train and test our machine learning model. Table \ref{table_features} describes the features derived from our input data. 

\begin{table}[!t]
\renewcommand{\arraystretch}{2.4}
\caption{List Of The Extracted Features}
\label{table_features}
\centering
\begin{tabular}{c c c}
\dtoprule
No. & Description & Equation\\
\hline
1 & \makecell{Maximum of Swing angle, \\ Gait Acceleration and Gait velocity}  & $max(s_{1},s_{2},...,s_{N})$\\
\hline
2 & \makecell{Mean of Swing angle, \\ Gait Acceleration and Gait velocity } & $\mu_{s} = \frac{1}{N}\sum_{i=1}^{N}s_{i}$\\
\hline
3 & \makecell{Standard deviation of Swing angle, \\ Gait Acceleration and Gait velocity } & $\sigma_{s} = \sqrt{\frac{1}{N}\sum_{i=1}^{N} (s_{i} - \mu_{s}) } $\\
\hline
4 & \makecell{RMS of Swing angle, \\ Gait Acceleration and Gait velocity}  & $x_{RMS} = \sqrt{\frac{1}{N}\sum_{i=1}^{N}s_{i}^{2}}$\\
\hline
5 & \makecell{Swing Period of stride $K$ \\ for non-bowling leg} & $sw = t(IC_{k}) - t(IS_{k})$\\
\hline
6 & \makecell{Stance Period of stride $K$ \\ for non-bowling leg} & $st = t(IS_{k+1}) - t(IC_{k})$\\
\hline
7 &  \makecell{Swing stance ratio of stride $K$ \\ for non-bowling leg} & $sw/st$\\
\dbottomrule
\multicolumn{3}{c}{\footnotesize IC, IS are Initial Contact, Initial Swing, respectively}
\end{tabular}
\end{table}

In our approach, we used a Support Vector Machine (SVM) \cite{cortes1995support} model to classify the skill level of athletes. SVM is a supervised machine learning method that can solve binary classification problems. In order to apply this method to our system, we decomposed the multi-class skill assessment task into multiple binary classification sub-problems using the one-versus-one technique \cite{weston1998multi}. For a problem with $C$ classes, $\frac{C(C-1)}{2}$ classifiers are constructed where each classifier is trained on data from two classes. In our case, $C=3$ therefore three SVMs are trained. To assign the label of a test case at prediction time, a voting scheme is applied. SVM aims to find the optimal separating hyperplane that maximizes the margin between the two classes.
Let $T=\{(x_{1},y_{1}),(x_{2},y_{2}),...,(x_{n},y_{n})\}$ be a training dataset of $n$ points, where $x_{i}\in \mathbf{R}^{n}$ is the data features and $y_{i}\in \{-1,+1\}$ indicates the class of $x_{i}$. To find the maximum margin hyperplane we solved the optimization problem $\underset{\omega,b}{min} \frac{1}{2}||\omega||^2$ subject to $y_{i}(\omega \cdot x_{i} - b) \leq 1$, where the vector $\omega$ and the displacement $b$ can describe the hyperplane by the linear equation $y = \omega^T X + b$. We convert the problem of finding the optimal hyperplane into the equivalent \emph{Lagrangian dual problem} which is defined as:

\begin{align} \label{eq:7}
& \underset{\lambda}{maximize} \quad \sum^{n}_{i=1}\lambda_{i} - \frac{1}{2}\sum^{n}_{i=1}\sum^{n}_{j=1}\lambda_{i}\lambda_{j}y_{i}y_{j}\phi(x_{i},x_{j}) \\ \nonumber
& s.t \quad \sum_{i=1}^{n}\lambda_{i}y_{i} = 0 \\ \nonumber
& and \quad C \geq a_{i} \geq 0 \ \forall i \in [1,n] \nonumber
\end{align}

where $\lambda$ represents the Lagrangian multipliers vector, $C$ is the penalty constant parameter which is used to find a trade-off between generalization and error, and $\phi(x_{i},x_{j})$ is the kernel function of the SVM. The quadratic optimization problem in (\ref{eq:7}) can be solved with the \emph{Sequential Minimal Optimization} (SMO) method \cite{platt1998sequential}. Using the Lagrangian multipliers obtained by the SMO method we can find the optimized hyperplane. Finally, the decision function for classification is defined as:

\begin{equation}
    f(X) = sign(\sum_{i=1}^{m} \lambda_{i}y_{i}\phi(x_{i},x_{j}) - b)
\end{equation}

where $X$ is the test dataset of size $m$ containing bowling throws of the athlete. Then an unknown bowling throw $X$ is classified as follows:

\begin{equation}
    X \in \begin{cases}
                \textrm{Class A} \quad \textrm{if } f(x) > 0, \\
                \textrm{Class B} \quad \textrm{otherwise.}
            \end{cases}
\end{equation}

Details of the training process, the hyper-parameter tuning, and the overall classification performance of our model are discussed on Section V. 

\section{Evaluation}

In this section, we evaluated the proposed system in the processes of quality assessment, error detection and skill level assessment. 

\subsection{Experimental Setup}

In order to test our system in a real bowling environment, we conducted a proof-of-concept experiment at Pin City Bowling center in Thessaloniki, using athletes of the 'Deka Korynes' Bowling Club. We recruited 9 right-handed bowlers (weight: 55-90 kg, height: 165-188 cm, age: 20-67), including 3 novice, 3 intermediate and 3 expert athletes. Expert bowlers had more than five years of experience and had participated in at least one official international tournament. Intermediate bowlers had 2-5 years of practice bowling and no participation in official tournaments. Novice bowlers had at most one year of practice bowling and no participation in tournaments. In addition, we recruited two professional coaches to provide templates that were used in quality assessment analysis. All athletes wore and configured the sensors according to our system instructions, and they were supervised by the two coaches. After the setup process and five warmup throws, each athlete performed 50 bowling throws. Bowling efforts that were considered to be illegal or committed a foul were excluded from our dataset. We obtained 450 throws ($9 \ \textrm{athletes} \times 50 \ \textrm{throws}$) using our Android application. Finally, for each throw of the process we captured a video of the bowler's motion. The videos were used to validate our system in the evaluation process.

\subsection{Quality Assessment Evaluation}

\begin{figure}[!t]
\centering
\includegraphics[width=3.5in]{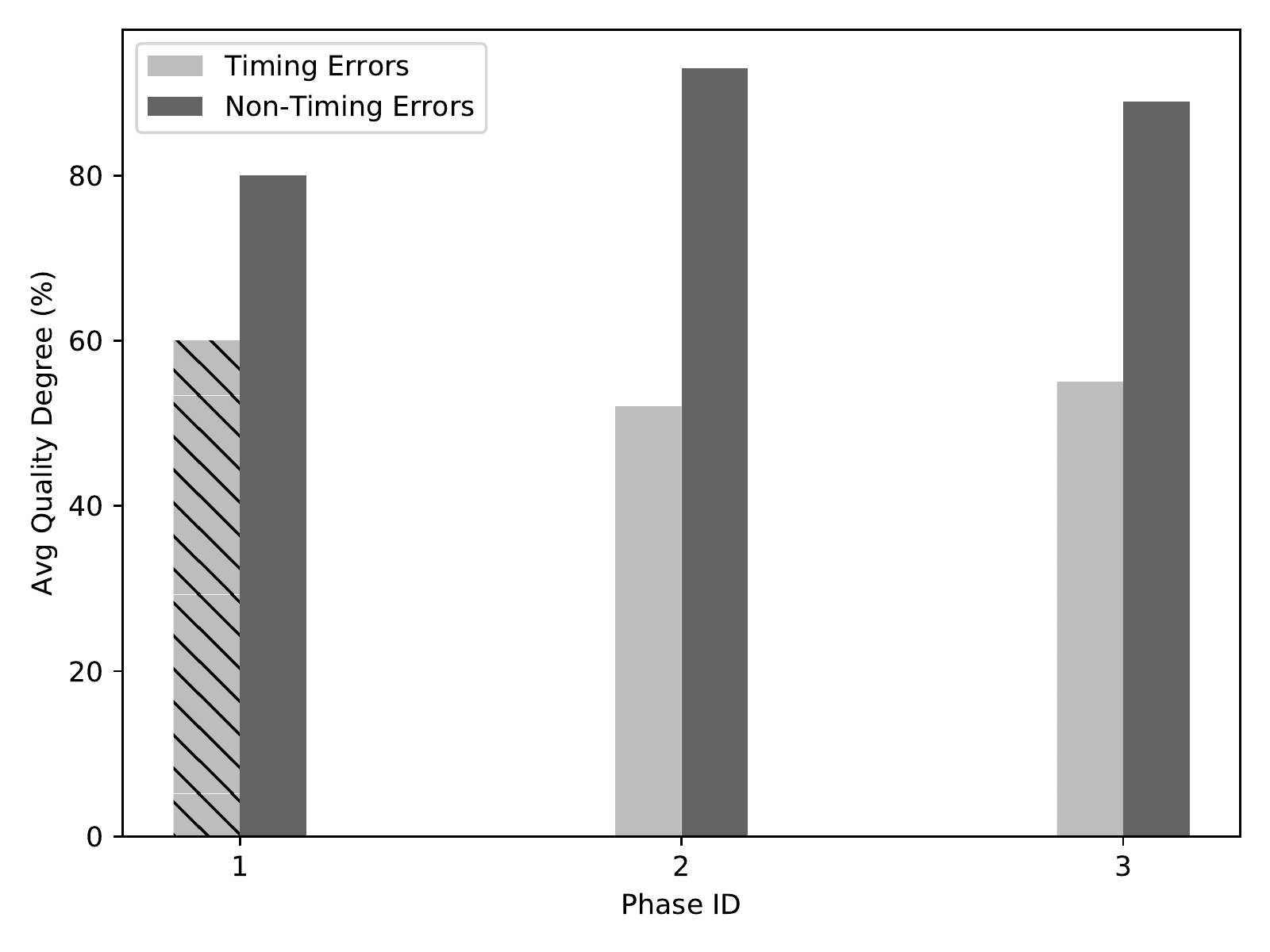}
\caption{ The comparison of average Quality Degree between data with timing errors and non-timing errors for the three phases.}
\label{fig_qd}
\end{figure}

To evaluate the accuracy of the quality assessment process, we asked the bowlers to perform 10 throws in which each one was as similar as possible to the motion of the coach. Then, the same bowlers performed 10 throws with evident timing errors for each significant bowling phase. The significant phases of the five-step approach are three, meaning that 30 throws are captured for each athlete. We applied the quality assessment process and measured the average quality degree of the collected data. The results of this metric are shown in Figure \ref{fig_qd}, where the average quality degree of each phase for timing error data and error-free data are compared. Statistical comparisons were done using paired T-test where p-values of $< 0.05$ were considered significant. There were significant statistical differences in the average quality degrees of throws with timing errors for phase 1 (M=60, SD=1.9), phase 2 (M=52, SD=2.1), phase 3 (M=55, SD=2.5) and their corresponding average quality degrees for throws with no timing errors for phase 1 (M=80, SD=1.8), phase 2 (M=93, SD=1.5), phase 3 (M=89, SD=1.8). Thus, the Quality Assessment analysis implemented in our system can be used to quantify the quality of the significant phases of a bowling throw.

In addition, we investigated the influence of the number of templates in our quality assessment approach. We used the 90 throws (10 throws per athlete) with no timing errors that were captured for the quality evaluation, and we compared the averaged quality degree for various number of templates. Using One-way ANOVA for each phase, we found that there were no statistical significant differences between the means of Quality degrees regarding the number of templates. This is presumably associated with the fact that the coaches can execute nearly similar throws, leading to highly correlated templates. Despite the weak influence in the Quality Degree metric, the number of templates can negatively impact the execution time of our quality assessment process. The minimum average execution time is 210ms for one template, and the maximum is 1100ms for six templates. Due to the weak influence of number of templates in Quality Degree and the computational power constraints of mobile devices, we employed only one template on our quality assessment process.

\begin{table}[!t]
\renewcommand{\arraystretch}{2.4}
\caption{Quality Degree and time execution for different number of templates.}
\label{table_QualityDegree}
\centering
\begin{tabular}{c ccc c}
\dtoprule
\multirow{3}{*}[3pt]{ \thead{Number of \\ Templates}}
    &   \multicolumn{3}{c}{\thead{Avg\\ Quality Degree (\%)}}
        &   \multirow{3}{*}{\thead{Avg\\ Execution \\ Time\\ (ms)}} \\
    \cmidrule(lr){2-4}
    & Phase 1 & Phase 2 & Phase 3 &     \\
    \midrule
1   & $80 \pm 4.56$ & $ 92 \pm 3.44 $ & $ 88 \pm 5.10 $ & 210 \\
2   & $80 \pm 2.33$ & $ 92 \pm 3.44 $ & $ 89 \pm 3.71 $ & 390 \\
3   & $81 \pm 1.90$ & $ 93 \pm 1.50 $ & $ 89 \pm 1.80 $ & 500 \\
4   & $81 \pm 1.90$ & $ 93 \pm 1.50 $ & $ 89 \pm 1.80 $ & 632 \\
5   & $83 \pm 1.78$ & $ 95 \pm 1.43 $ & $ 90 \pm 1.76 $ & 855 \\
6   & $83 \pm 1.78$ & $ 95 \pm 1.43 $ & $ 90 \pm 1.76 $ & 1100 \\
    \dbottomrule
\end{tabular}
\end{table}

\subsection{Common Error Detection Evaluation}

We evaluate the common error detection process using a subset of the data from quality assessment evaluation. The coaches watched the recorded high frame rate videos of the bowlers' throws to label the data with the corresponding common error type. Precision and recall metrics are used to evaluate our results obtained from \emph{Error Detection} process. Precision metric measures the ratio of the correctly detected errors over the total number of detected errors and it is described by the equation:
\begin{equation}\label{eq:10}
    Precision = \frac{TP}{TP + FP}
\end{equation}
Recall metric measures the ratio of the correctly detected errors over the total number of errors in our dataset and it is described as follows: 
\begin{equation}\label{eq:11}
    Recall = \frac{TP}{TP + FN}
\end{equation}
where $TP$, $FP$ and $FN$ are true positive, false positive and false negative, respectively. As shown in Table \ref{table_error_detection}, the averaged precision and recall over all types of errors are $90\%$ and $84\%$, respectively. Overall the results suggest that \emph{Common Error Detection} process can be used to identify the four bowlers errors.  

\begin{table}[!t]
\renewcommand{\arraystretch}{2.4}
\caption{Common Error Detection Results}
\label{table_error_detection}
\centering
\begin{tabular}{c c c}
\dtoprule
Type Error & Precision & Recall \\
\hline
1 & 0.88 & 0.75\\
2 & 0.89 & 0.77\\
3 & 0.94 & 0.90\\
4 & 0.90 & 0.95\\
\dbottomrule
\end{tabular}
\end{table}

\subsection{Performance Evaluation}

\begin{figure*}[!t]
\centering
\subfloat[]{\includegraphics[width=2.5in]{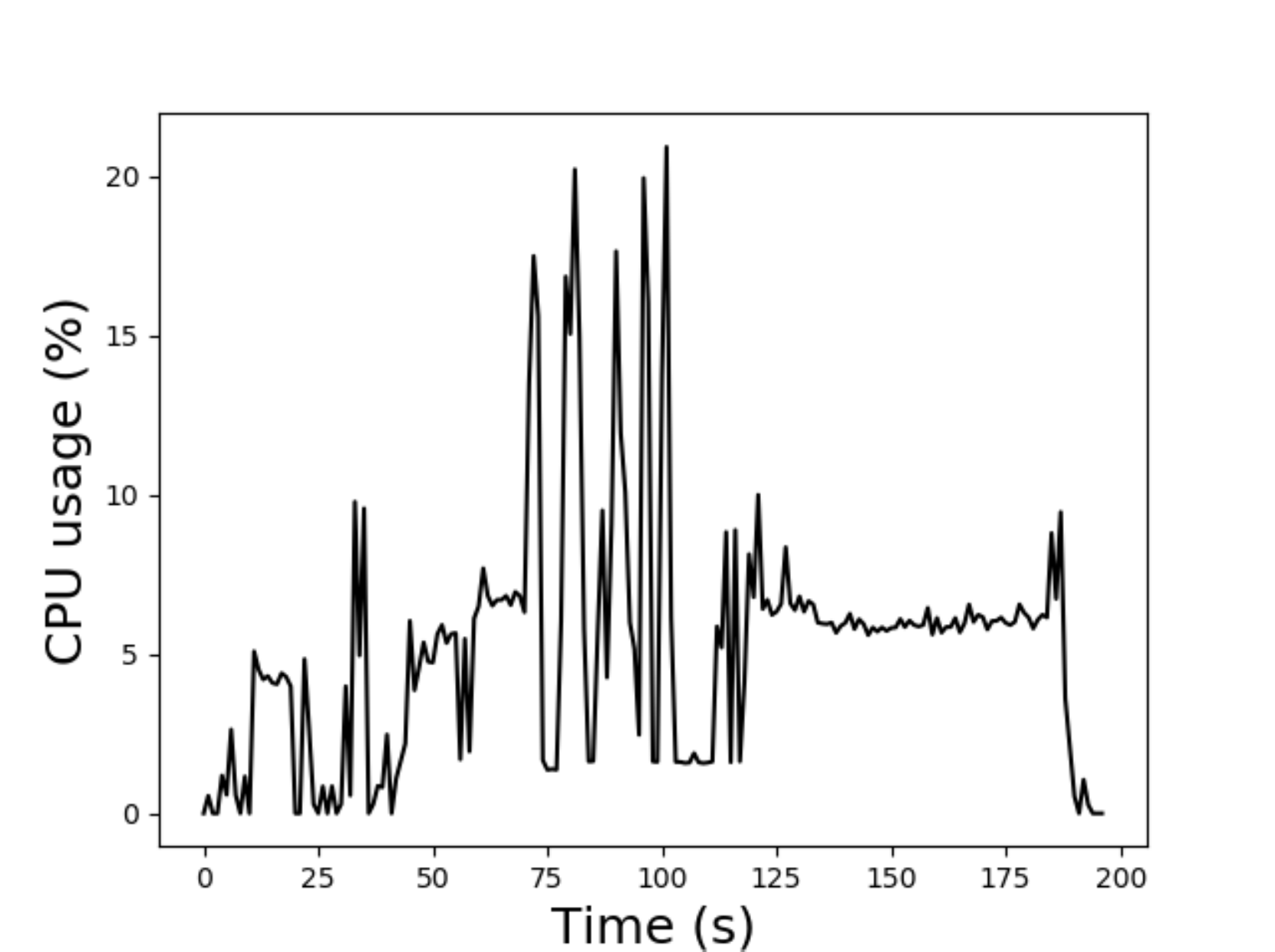}%
\label{fig_cpu}}
\hfil
\subfloat[]{\includegraphics[width=2.5in]{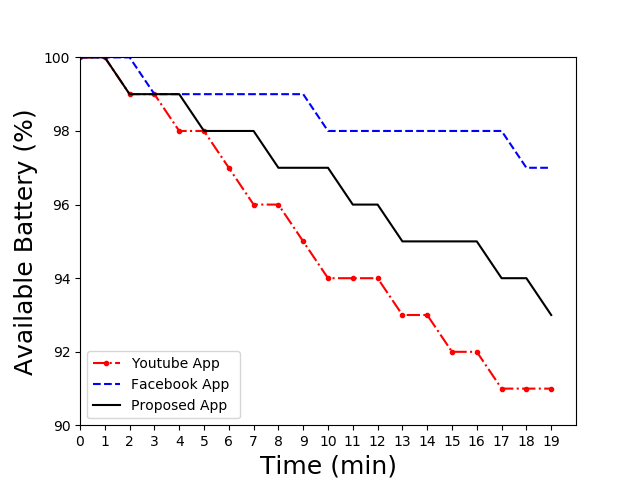}%
\label{fig_energy}}
\caption{CPU and energy performance of the proposed mobile application.}
\label{fig_performance}
\end{figure*}

Quality assessment process and error detection are both performed during the course of the game in athletes' mobile device. Mobile devices are characterized with limited process power and energy capacity. To assure adequate performance of our mobile application, we evaluate the CPU and energy consumption using a Xiaomi Redmi A2 (Snapdragon 660, Octa-Core Cryo 2.2 GHz, battery 3010 mAh) smartphone.

We captured the CPU usage of our Android application during the sensors configuration, the quality assessment, and the error detection of a single bowler's throw. During this process (approximate 200 seconds), there was no other application or task running in the background of the Android device. The results are shown in Figure \ref{fig_cpu}. The maximum CPU usage is $23\%$ and the average usage is $5.23\%$. Comparing our results with studies investigating the CPU usage of Android applications \cite{DORFER2020189} \cite{LINARESVASQUEZ20171}, we see that our mobile application has an acceptable CPU usage.

To investigate the energy consumption, we captured the battery level of the mobile device during a session with 15 throws. This session lasted about 20 minutes, during which an athlete used the mobile application to configure the sensors and get visual feedback for each throw. At the start of the experiment, the mobile device was fully charged and no other applications were running on the background. To capture the battery data, Android Debug Bridge (ADB) and \emph{dumpsys} tools were used. Figure \ref{fig_energy} shows the energy consumption during the session. The X axis is time in minutes, and the Y axis is the percentage of the available battery of the mobile device. The available battery of the device running the proposed application is decreasing from 100\% to 93\% after a session of 20 minutes. We contrasted the energy consumption of the proposed application with two popular mobile applications, Facebook and Youtube. The two applications were tested in the same mobile device under the same testing conditions. As we can see from figure \ref{fig_energy}, the total energy consumption of the proposed application is similar to the two popular applications.

\subsection{Skill Assessment Evaluation}

We used the 450 datasets obtained from 9 athletes to built our skill assessment machine learning model. The  data collection was split into a training set (80\% of the dataset) and a test set (20\% of the dataset). The training set (360 throws) is used for the training process and the test set (90 throws) is used for testing the machine learning model. To avoid overfitting of our model, we used 10-fold cross validation. We obtained the optimal hyperparameters of the SVMs classifiers using the grid search tuning method. RBF kernel function was found to give the best results for our model. The hyperparameters of the SVM with RBF kernel are $C$ and $\gamma$, which were validated in the ranges of $[10^{-1}, 10^{4}]$ and $[10^{-6}, 10^{-1}]$, respectively. Macro averaged F1 metric is defined as the score function of our grid search method. The F1-score can be used for multiclass classification problems and is defined as:

\begin{equation}
    F1 = 2 \cdot \frac{precision \cdot recall}{precision + recall}
\end{equation}

where precision and recall are defined by equations (\ref{eq:10}) and (\ref{eq:11}), respectively. We obtained the best results for macro averaged F1 metric when $C=1$ and $\gamma=10^{-6}$. The classification results of our trained model are shown in Table \ref{table_svm_results}.

\begin{table}[!t]
\renewcommand{\arraystretch}{2.4}
\caption{Classification results of the proposed model}
\label{table_svm_results}
\centering
\begin{tabular}{c c c c}
\dtoprule
Skill & Precision & Recall & F1 \\
\hline
Expert & 0.88 & 0.75 & 0.75\\
Intermediate & 0.89 & 0.77 & 0.75\\
Novice & 0.94 & 0.90 & 0.75\\
\hline
Weighted Avg  & 0.94 & 0.90 & 0.75  \\
\dbottomrule
\end{tabular}
\end{table}

\subsection{Discussion}

The proposed system can effectively and efficiently assess the quality of bowlers' throws. The results of the experiments suggest that our method can accurately quantify the characteristics of the significant phases of a bowling throw. In addition, we shown that the number of templates has low influence on the quality assessment process.

The proposed method for detecting common bowling errors can accurately identify four types of motion errors, with an averaged precision of 90\% and recall 84\%. The method is based on comparison of statistical parameters between motion data of athletes and templates. For this reason, the method requires a range predefined by the coaches to detect motion errors. Therefore, the common error detection process is limited by the predefined range.

The system after each throw provides real-time feedback with regard to the quality of the throw and the detected errors. This feedback can help bowlers and coaches track their motion data and assess their performance. Thus, the processes of quality assessment and error detection need to perform efficiently on user mobile devices. The results from the performance evaluation suggest that the mobile application (running the two processes) have an acceptable CPU usage and energy consumption.

Regarding the skill assessment process, the results of our experiments indicate that the proposed cloud-based system can accurately recognize the skill level using motion characteristics such as swing angle, gait velocity, etc. The SVM model achieved an average precision of 94\% and recall of 90\%, showing that it can effectively identify the skill level of an athlete. Moreover, the classification results suggest that novice bowlers have different motion characteristics from those of experts and intermediates resulting better recognition performance for novice bowlers.

However, the proposed system has a number of limitations. The system is tested and evaluated with motion data from right-handed bowlers that performed the five step approach. Therefore, the current version of our system is not suitable for motion techniques that require more than five steps. One possible limitation of our work is the size of the captured sensing data. Our dataset was limited to data captured from a small group of athletes (9 right-handed bowlers). Capturing more data will allow us to extend our research to other bowling-techniques and game styles (e.g. the four step approach). For that reason it necessary to gather more data from a wide range of skill level and age groups. Moreover, in order for the system to work, proper initialization of the baseline (see section V.A) is needed. Wrong baseline initialization or inconsistent and unexpected arm motion (for example moving the arm out of the sagittal plane) will result in misleading conclusions and wrong performance evaluation. 

We believe that Wearable Computing and computer science will have a profound impact on sport science. Our future work will aim to improve the overall performance and extend the current system to work with other motion techniques. Our approach can also be applicable to other target sports and physical activities (e.g. shot put). Thus, we aim to collect data and extend the proposed system for other sports. Finally, we envision that by providing athletes with sensors in other human body positions, the proposed system can be extended to assess and monitor the performance of other throwing sports.

\section{Conclusion}

Bowler's motion technique is an important factor in delivering a successful bowling throw into the alley. Motion characteristics during a throw, such as gait and arm synchronization, maximum angle of the arm, and gait speed, are used by the coaches to provide feedback to bowling athletes. In this study, we proposed a novel system for automating the coaching procedure of ten-pin bowling.  The proposed approach includes two inertial measurement units, a mobile application, and a cloud based server. Motion data obtained by the bowler's sensors on non-bowling leg and bowling wrist were used for assessing the quality of the throw, identifying common bowling errors, and recognizing the skill level of the bowler. The streaming motion data are sent through Bluetooth Low Energy on a mobile device, which in turn performs quality assessment and error detection analysis. The system analyzes the data and displays the results in real-time on the mobile device. Moreover, the user can upload their data on a designated cloud server where the system assesses the skill level of the bowler using a machine learning model. Based on the results of our experiments with bowlers' motion data, our system can assess the quality of the throw and identify common errors with good performance. The results also shown that our SVM classifier achieves high accuracy in bowler's skill assessment.


\bibliographystyle{ACM-Reference-Format}
\bibliography{sample-base}

\newpage

\section{List Of Changes}

\begin{itemize}

\item Improve related work. Include literature review of wearable computing and IMUs in other sports.
\item Based on the related work and on the reviewers suggestions we strengthen the discussion section including the limitations pointed out by the reviewers. We also justify the technical novelty of our system based on the related work.
\item  We reduced the length of the paper by moving the platform architecture and user interface details to a separate supplementary pdf document.
\item Fixing minor grammar errors.
\end{itemize}


\end{document}